# Excitation of Coherent States: Wave Function Development and Analysis

Gaurav Singh, Anshul Choudhary
Netaji Subhas Institute of Technology, Delhi University, Delhi, India
And Prof. T.R.Sheshadri
Department of Astrophysics and Astronomy, Delhi University, Delhi, India

**ABSTARCT**

Agarwal and Tara, in 1991 introduced a new class of states defined as 'm' times application of creation operator to Coherent States known as Excited Coherent States (ECS) or Photon Added Coherent States (PACS). They are neither completely quantum nor completely classical. Here we present and develop these Excited Coherent Sates from a basic and more approachable Wave-function approach. We have derived the ECS wave function as a blend of Coherent States and Fock States and thus established them as a result of Quantum fluctuations (represented by Fock states) on Coherent States. We further derived and analyzed basic relations such as wave packet width and uncertainty relation in a more generalized form and presented their development with time. Another important property of ECS is Quadrature Squeezing. Here we also present a general analysis of squeezing in ECS and derived conditions on parameters for squeezing.

## 1. Introduction

While studying quantum mechanics, a natural question which arises is how to obtain the classical limit of a quantum system. There are several approaches to address this question. One of the approaches is by constructing states called coherent states. These are near classical states in the sense that the quantum uncertainty is the minimum possible and the mean value of the position operator follows the classical equations of motion. If the uncertainty could be made to zero the probability density function constructed from these states would be a delta function and the mean value would have been the classical value of the position. Then the fact that the mean value follows the classical equation of motion would imply that the particle position (defined uniquely in such a case) would follow the classical equations of motion. However, the uncertainty cannot be made to zero and hence, the best we can have as the classical limit is to have the minimum uncertainty state. In this sense coherent states are the classical limit of a quantum system. These states were originally realized by Erwin Schrödinger in 1926 while searching for solutions of his proposed equation (Schrödinger Equation) that resemble classical motion. In 1963, the proper theoretical framework was given by Roy.J.Glauber and Sudershan [1].

## 2. Coherent States

Coherent states are Eigen states of the annihilation operator $\hat{a}$. Denoting the Eigen value by $\alpha$ and the corresponding Eigen state by $|\alpha\rangle$, we have, $\hat{a}|\alpha\rangle = \alpha|\alpha\rangle$

Since a general coherent state is not the ground state and since the eigenstates $|n\rangle$ form a complete set, we can express the state $|\alpha\rangle$ as a combination of $|n\rangle$'s. The precise connection happens to be,

$$|\alpha\rangle = \exp\left(\tfrac{1}{2}|\alpha|^2\right) \sum_{n=0}^{\infty} \frac{\alpha^n}{\sqrt{n!}}|n\rangle \tag{2.1}$$

Consider the special case when $\alpha = 0$. We immediately see that this ket obeys the same Eigen value equation as that by the ground state. Thus the ground state is simply a special case of coherent states. Using $\hat{a}^{\dagger}|n\rangle = \sqrt{n+1}|n+1\rangle$, we have for n = 0, $\hat{a}^{\dagger}|0\rangle = |1\rangle$

Further, by operating $\hat{a}^{\dagger}$ on $|0\rangle$, n times we have,

$$|n\rangle = \frac{\left[\hat{a}^{\dagger}\right]^n}{\sqrt{n!}}|0\rangle \tag{2.2}$$

Therefore, higher (excited) energy excited states can be constructed by successive application of the creation operator, $\hat{a}^{\dagger}$ on the ground state. A natural question arises as to the effect of the creation operator on a general coherent state. These are called the Excited Coherent States and has been studied by Agarwal and Tara in 1991[2], and are defined by

$$|z, m\rangle = k_m \left[\hat{a}^{\dagger}\right]^m |\alpha\rangle \tag{2.3}$$

where $k_m = \dfrac{1}{\sqrt{N_m m!}}$ =Normalization constant &

$$N_m = \sum_{r=0}^{m} {}^mC_r \cdot \frac{\beta^r}{r!} = L_m(-|\alpha|^2)$$

where $L_m$ is the Laguerre polynomial of order 'm '[3] . Also now onwards we will take, $\beta = |\alpha|^2$ for brevity.

### 3. Building ECS Wave function

Using equation (2.1) and (2.3)) the excited state of order 'm', $|z;m\rangle$ becomes

$$|z;m\rangle = \frac{1}{\sqrt{N_m m!}} \exp\left(-\frac{|\alpha|^2}{2}\right) \sum_{n=0}^{\infty} \frac{\alpha^n}{\sqrt{n!}} \sqrt{(n+1)(n+2)(n+3)\cdots(n+m)}\,|n+m\rangle \tag{3.1}$$

Also we know the coordinate representation of Fock states as [4],

$$\langle x\,|\,n\rangle = \psi_n(x) = N_n H_n(\varepsilon) \exp\left(-\frac{\varepsilon^2}{2}\right) \tag{3.2}$$

So,

$$\langle x\,|\,n+m\rangle = \psi_{n+m}(x) = N_{m+n} H_{n+m}(\varepsilon) \exp\left(-\frac{\varepsilon^2}{2}\right)$$

$$= \left[\frac{\gamma}{2^{n+m}(n+m)!\sqrt{\pi}}\right]^{\frac{1}{2}} H_{n+m}(\varepsilon) \exp\left(-\frac{\varepsilon^2}{2}\right)$$

Substituting this in (3.1), we can calculate the wave function for m<sup>th</sup> order excited coherent state as

$$\psi_{z,m}(x) = \frac{1}{2^{\frac{m}{2}}\sqrt{N_m m!}} \left(\frac{\gamma}{\sqrt{\pi}}\right)^{\frac{1}{2}} \exp\left(-\frac{\varepsilon_0^2}{4} - \frac{\varepsilon^2}{2}\right) \sum_{n=0}^{\infty} \left(\frac{\varepsilon_0}{2}\right)^n \frac{1}{n!} H_{n+m}(\varepsilon) \tag{3.3}$$

Now an interesting problem arises in above eq, we know that the generating function of hermite polynomial is given by $\exp(-s^2 + 2s\varepsilon) = \sum_{n=0}^{\infty} \frac{s^n}{n!} H_n(\varepsilon)$ [5]; same as in our equation with **s** replaced by $\frac{\varepsilon_0}{2}$ but we have $H_{n+m}$ instead of $H_n$ so we need to break $H_{n+m}$ and it could be easily shown that

$$\sum_{n=0}^{\infty}\left(\frac{\varepsilon_0}{2}\right)^n \frac{1}{n!} H_{n+m}(\varepsilon) = \sum_{n=0}^{\infty}\left(\frac{\varepsilon_0}{2}\right)^n \frac{1}{n!} H_n(\varepsilon) H_m\left(\varepsilon - \frac{\varepsilon_0}{2}\right) \qquad (3.4)$$

**(*See Appendix for the Complete Proof*)**

Now, using above the wave function becomes,

$$\psi_{z,m}(x) = \frac{1}{2^{\frac{m}{2}}\sqrt{N_m m!}}\left(\frac{\gamma}{\sqrt{\pi}}\right)^{\frac{1}{2}} \exp\left(-\frac{\varepsilon_0{}^2}{4} - \frac{\varepsilon^2}{2}\right)\sum_{n=0}^{\infty}\left(\frac{\varepsilon_0}{2}\right)^n \frac{1}{n!} H_n(\varepsilon) H_m\left(\varepsilon - \frac{\varepsilon_0}{2}\right) \qquad (3.5)$$

and substituting, $\exp\left(-\frac{\varepsilon_0{}^2}{4} + \varepsilon\varepsilon_0\right) = \sum_{n=0}^{\infty}\left(\frac{\varepsilon_0}{2}\right)^n \frac{1}{n!} H_n(\varepsilon)$ ,in above

we get,

$$\psi_{z,m}(x) = \frac{1}{2^{\frac{m}{2}}\sqrt{N_m m!}}\left(\frac{\gamma}{\sqrt{\pi}}\right)^{\frac{1}{2}} \exp\left(-\frac{1}{2}(\varepsilon - \varepsilon_0)^2\right) H_m\left(\varepsilon - \frac{\varepsilon_0}{2}\right) \qquad (3.6)$$

Also we know that coherent state wave function [6] is given by:

$$\psi_\alpha(x) = \left(\frac{\gamma}{\sqrt{\pi}}\right)^{\frac{1}{2}} \exp\left(-\frac{1}{2}(\varepsilon - \varepsilon_0)^2\right) \qquad (3.7)$$

Substituting this expression in equation (3.6), we get

$$\psi_{z,m}(x) = \frac{1}{2^{\frac{m}{2}}\sqrt{N_m m!}} H_m\left(\varepsilon - \frac{\varepsilon_0}{2}\right)\psi_\alpha(x) \qquad (3.8)$$

$\psi_{z,m}(x)$ = wave-function for m$^{\text{th}}$ order excitation of coherent state.

Now an interesting observation:

Put **m = 0** in eq (3.8), we get coherent states i.e. $\psi_{z,0}(x) = \psi_\alpha(x)$

Put $\alpha$ **=0** in eq (3.8), we get Fock states or eigenfunctions of oscillator i.e.

$\psi_{z,0}(x) = \psi_m(x)$

As we can see normal coherent states are nothing but ground states(m=0) of more general **Excited Coherent States** Thus Excited Coherent states represent or exhibit mixtures of both coherent states (which are Quantum mechanical analogs of classical oscillator ) and Fock states (strictly quantum with no classical analog). Thus they **represent Quantum fluctuations in simple quantum coherent states.**

## 4. Time Dependent Wave-function

The time-dependent wave function [7] can be calculated as

$$\left|\psi(x,t)\right\rangle = \sum_{n=0}^{\infty} C_n \exp\left(-\frac{iE_n t}{n}\right)\left|n\right\rangle$$

$$= \sum_{n=0}^{\infty} C_n \exp\left(-i\left(n+\tfrac{1}{2}\right)\omega t\right)\left|n\right\rangle \tag{4.1}$$

So, substituting $\left|\psi_{z,m}(x)\right\rangle = \sum_{n=0}^{\infty} C_n \left|n\right\rangle$ from equation (3.3) in above equation, we get

$$\psi_{z,m}(x,t) = \frac{1}{2^{\frac{m}{2}}\sqrt{N_m m!}}\left(\frac{\gamma}{\sqrt{\pi}}\right)^{\frac{1}{2}} \exp(-i\omega t)\exp\left(-\frac{\varepsilon_0{}^2}{4}-\frac{\varepsilon^2}{2}\right)\sum_{n=0}^{\infty}\left(\frac{\varepsilon_0}{2}\exp(-i\omega t)\right)^n \frac{1}{n!} H_{n+m}(\varepsilon) \tag{4.2}$$

As done previously in the case of eq (3.3), here also

$$\sum_{n=0}^{\infty}\left(\frac{\varepsilon_0}{2}\exp(-i\omega t)\right)^n \frac{H_{n+m}(\varepsilon)}{n!} = \left(\sum_{n=0}^{\infty}\left(\frac{\varepsilon_0}{2}\exp(-i\omega t)\right)^n \frac{H_n(\varepsilon)}{n!}\right) H_m\left(\varepsilon - \frac{\varepsilon_0}{2}\exp(-i\omega t)\right)$$

Again using

$\exp\left(-s^2 + 2s\varepsilon\right) = \sum_{n=0}^{\infty} \frac{s^n}{n!} H_n(\varepsilon)$, where $s = \frac{\varepsilon_0}{2}\exp(-i\omega t)$

Equation (4.2) can be re-written as

$$\psi_{z,m}(x,t) = \frac{1}{2^{\frac{m}{2}}\sqrt{N_m m!}}\left(\frac{\gamma}{\sqrt{\pi}}\right)^{\frac{1}{2}} \exp\left(-\frac{\varepsilon_0{}^2}{4}\left(1+\exp(-2i\omega t)\right)-\frac{\varepsilon^2}{2}+\varepsilon\varepsilon_0\exp(-i\omega t)-\frac{i\omega t}{2}\right) H_m\left(\varepsilon-\frac{\varepsilon_0}{2}\exp(-i\omega t)\right) \tag{4.3}$$

Now we know that the time dependent Coherent state wave function [8] is

$$\psi_{z,m}(x)=\left(\frac{\gamma}{\sqrt{\pi}}\right)^{\frac{1}{2}}\exp\left(-\frac{\varepsilon_0{}^2}{4}\left(1+\exp(-2i\omega t)\right)-\frac{\varepsilon^2}{2}+\varepsilon\varepsilon_0\exp(-i\omega t)-\frac{i\omega t}{2}\right) \tag{4.4}$$

So, now Equation (4.3) becomes

$$\psi_z(x,t)=\frac{1}{2^{\frac{m}{2}}\sqrt{N_m m!}}\psi_\alpha(x,t)H_m\left(\varepsilon-\frac{\varepsilon_0}{2}\exp(-i\omega t)\right) \tag{4.5}$$

Now Let us study the time evolution of first order ECS wave packet by putting **m=1** in eq (4.5) and finding the probability density function as

$$\left|\psi_{z,1}(x,t)\right|^2=\frac{2\gamma}{(1+\beta)\sqrt{\pi}}\exp\left(-(\varepsilon-\varepsilon_0\cos\omega t)^2\right)\left(\varepsilon^2-\varepsilon\varepsilon_0\cos\omega t+\frac{\varepsilon_0{}^2}{4}\right) \tag{4.6}$$

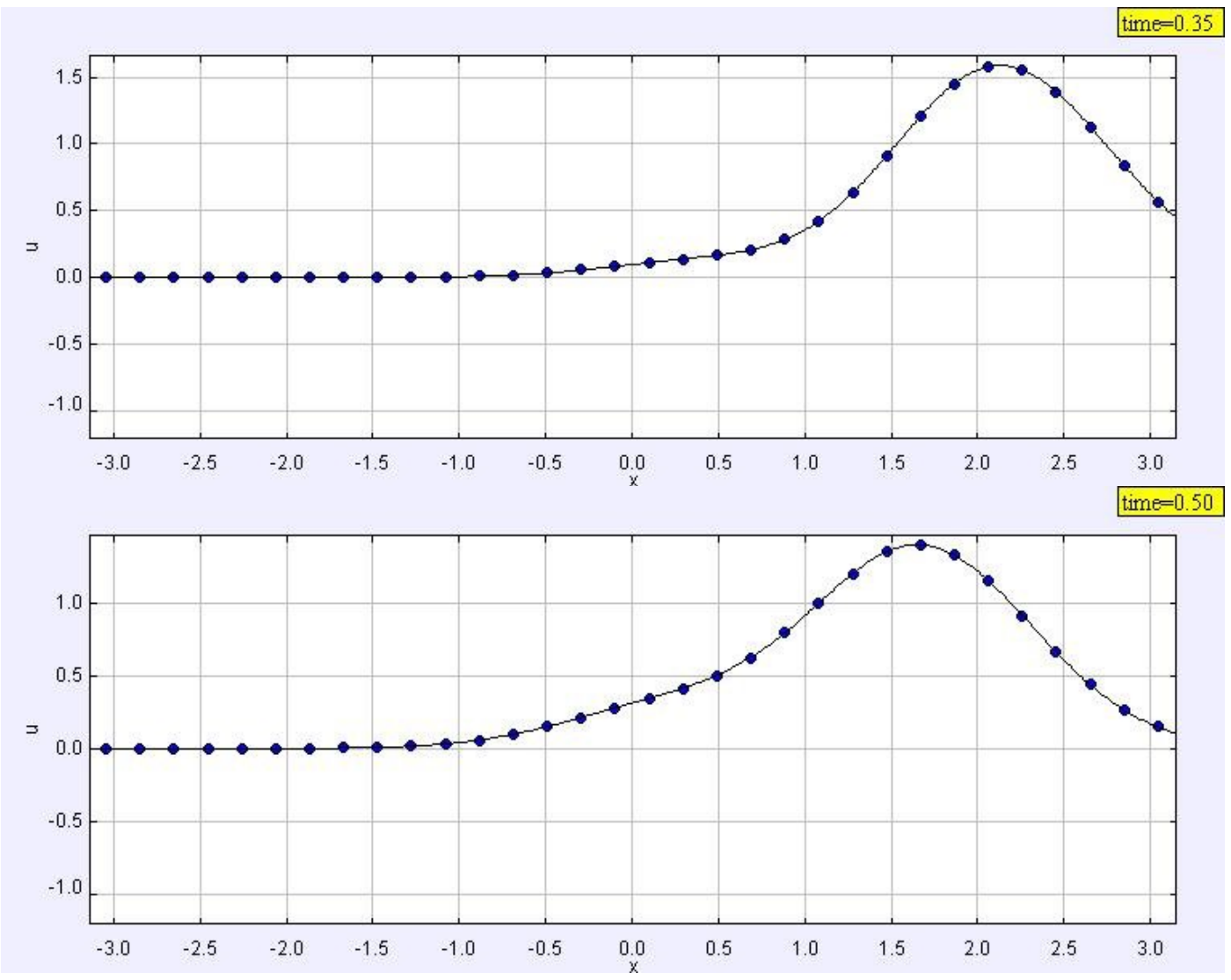

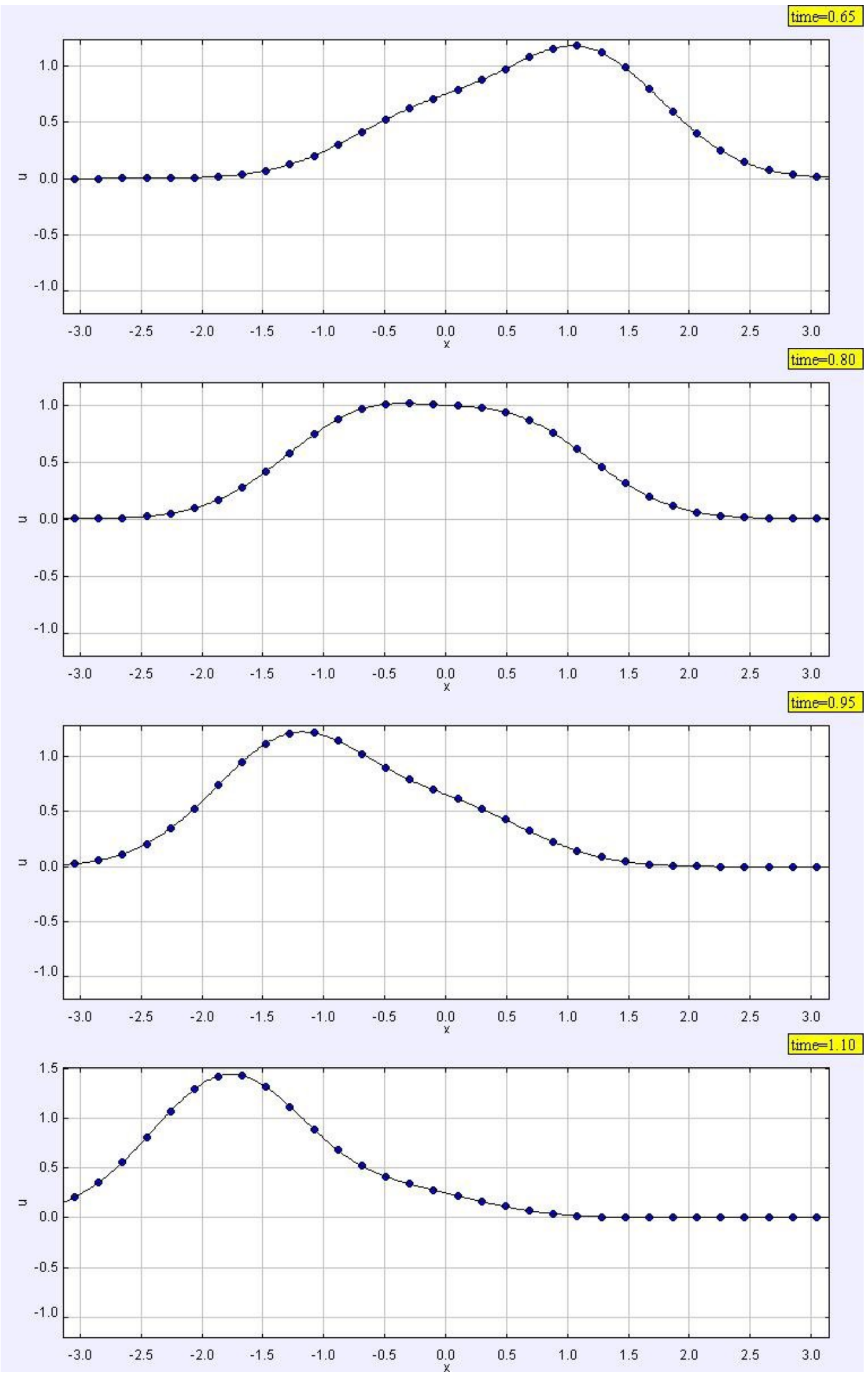

**Fig: 1 (Time Evolution of Wave packet):** The **Y-axis** denotes probability density function and **X-axis** denotes $\varepsilon$ **.**As the snapshots suggest, the wave packet which was initially Gaussian, gets distorted with time. These distortions or fluctuations are nothing but quantum excitations in coherent states

## 5. Some Properties of the Wave-function

Since, we can represent position $\hat{x}$ and momentum $\hat{p}$ operators in terms of creation ( $\hat{a}$ ) and annihilation ( $\hat{a}^{\dagger}$ ) operators as

$$\hat{x} = \sqrt{\frac{\hbar}{2\mu\omega}}\left(\hat{a} + \hat{a}^{\dagger}\right) , \quad \hat{p} = \frac{1}{i}\sqrt{\frac{\mu\hbar\omega}{2}}\left(\hat{a} - \hat{a}^{\dagger}\right)$$

Now, $|z;m\rangle = |\psi_z(t=0)\rangle$

$$|\psi_z(t)\rangle = \frac{1}{\sqrt{N_m m!}}\exp\left(-\frac{|\alpha|^2}{2}\right)\sum_{n=0}^{\infty}\frac{\alpha^n}{\sqrt{n!}}\sqrt{(n+1)(n+2)(n+3)\cdots(n+m)}\exp\left(-i\left(n+\frac{1}{2}\right)\omega t\right)|n+m\rangle \tag{5.1}$$

The expectation value of position $\hat{x}_m$ is

$$\langle\hat{x}_m\rangle = \sqrt{\frac{\hbar}{2\mu\omega}}\langle\psi_z(t)\,|\,\hat{a}+\hat{a}^{\dagger}\,|\,\psi_z(t)\rangle \tag{5.2}$$

which could be easily calculated and finally we will get

$$\langle\hat{x}_m\rangle = \sqrt{\frac{2\hbar}{\mu\omega}}\frac{|\alpha|}{N_m}S_1\cos(\omega t-\phi) \tag{5.3}$$

where $S_1$ is a constant given by $S_1 = \frac{1}{m!}\sum_{r=0}^{m} {}^{m}C_r.\beta^r.\frac{(m+1)!}{(r+1)!}$ .Similarly, we can calculate $\langle\hat{x}_m^2\rangle$ as

$$\langle\hat{x}_m^2\rangle = \frac{\hbar}{2\mu\omega}\left[\frac{2S_2\beta}{N_m}\cos 2(\omega t-\phi)+\frac{S_3+S_4}{N_m}\right] \tag{5.4}$$

where $S_1$, $S_2$, $S_3$, $S_4$ are constants given by

$$S_2 = \frac{(m+2)!}{m!}\sum_{r=0}^{m} {}^{m}C_r.\beta^r.\frac{1}{(r+2)!}$$

$$S_3 = \frac{1}{m!}\sum_{r=0}^{m}\frac{{}^{m}C_r\beta^r m + {}^{m+1}C_{r+1}\beta^{r+1}}{r!}$$

$$S_4 = \frac{(m+1)!}{m!}\sum_{r=0}^{m} {}^{m+1}C_r.\beta^r.\frac{1}{(r)!}$$

Now, wave packet width is given by $\Delta x = \left[\left\langle x^2 \right\rangle - \langle x \rangle^2\right]^{\frac{1}{2}}$

So, substituting eq (5.2) and eq (5.4) in above , we get

$$\Rightarrow \boxed{\Delta x = \sqrt{\frac{\hbar}{2\mu\omega}}\left[\frac{4\beta}{N_m}\left(S_2 - \frac{S_1^{\,2}}{N_m}\right)\cos^2(\omega t - \phi) + \frac{S_3 + S_4 - 2S_2\beta}{N_m}\right]^{\frac{1}{2}}} \tag{5.5}$$

**Now $\Delta x$ varies with time and will be independent of time only if**

$S_2 - \frac{S_1^{\,2}}{N_m} = 0$ , which is clearly only for $m = 0$

$\Rightarrow$ So $\Delta x$ i.e. width of wavepacket is independent of time only for ground state coherent states.

Similarly

$$\boxed{\Delta p = \sqrt{\frac{\mu\omega\hbar}{2}}\left[\frac{4\beta}{N_m}\left(S_2 - \frac{S_1^{\,2}}{N_m}\right)\sin^2(\omega t - \phi) + \frac{S_3 + S_4 - 2S_2\beta}{N_m}\right]^{\frac{1}{2}}} \tag{5.6}$$

Let us call $\frac{4\beta}{N_m}\left(S_2 - \frac{S_1^{\,2}}{N_m}\right) = C_1$ and $\frac{S_3 + S_4 - 2S_2\beta}{N_m} = C_2$

so uncertainity relation $\Delta x.\Delta p$ would become

$$\boxed{\Delta x.\Delta p = \frac{\hbar}{2}\left[C_2^{\,2} + C_1C_2 + \frac{C_1^{\,2}}{4}\sin^2 2(\omega t - \phi)\right]^{\frac{1}{2}}} \tag{5.7}$$

Lets us put **m**=1 in Eq (5.5), i.e. first order Excited coherent states
We can easily find out

$$\Delta x_1 = \sqrt{\frac{\hbar}{2\mu\omega}}\left[\frac{\beta+3}{\beta+1} - \frac{4\beta}{(1+\beta)^2}\cos^2(\omega t - \phi)\right]^{\frac{1}{2}} \tag{5.8}$$

where as $\Delta x_0 = \sqrt{\frac{\hbar}{2\mu\omega}}$ is ground state width.

Also we can show that

$$\sqrt{\frac{\beta^{2}+3}{(\beta+1)^{2}}} \leq \frac{\Delta x_{1}}{\Delta x_{0}} \leq \sqrt{\frac{\beta+3}{\beta+1}}$$

Graphically this could be represented in Fig 2 and Fig 3 as

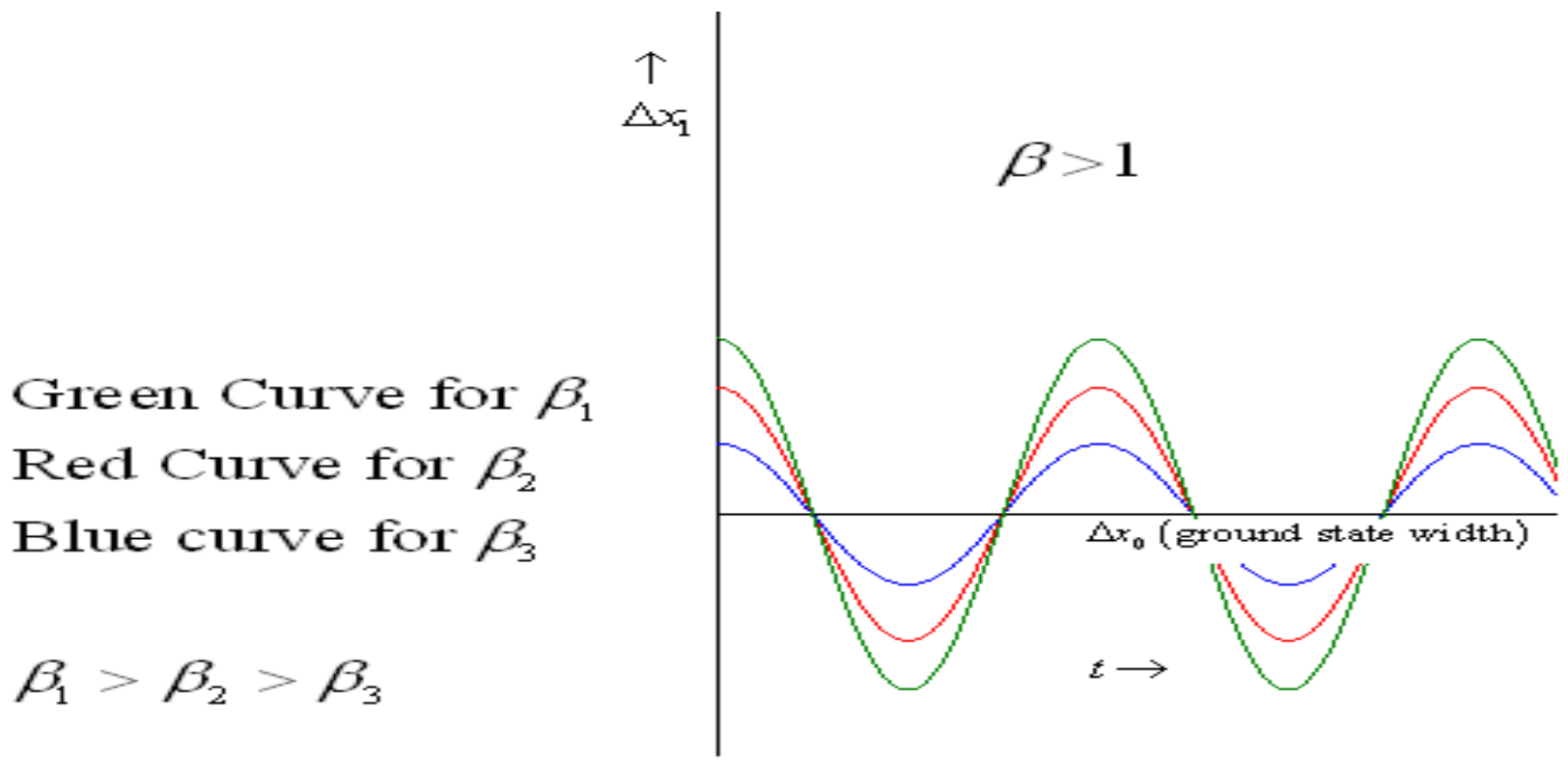

**Fig: 2** Variation of wave-packet width with time for different values of $\beta > 1$. As evident x-quadrature squeezing increases with increase in $\beta$.

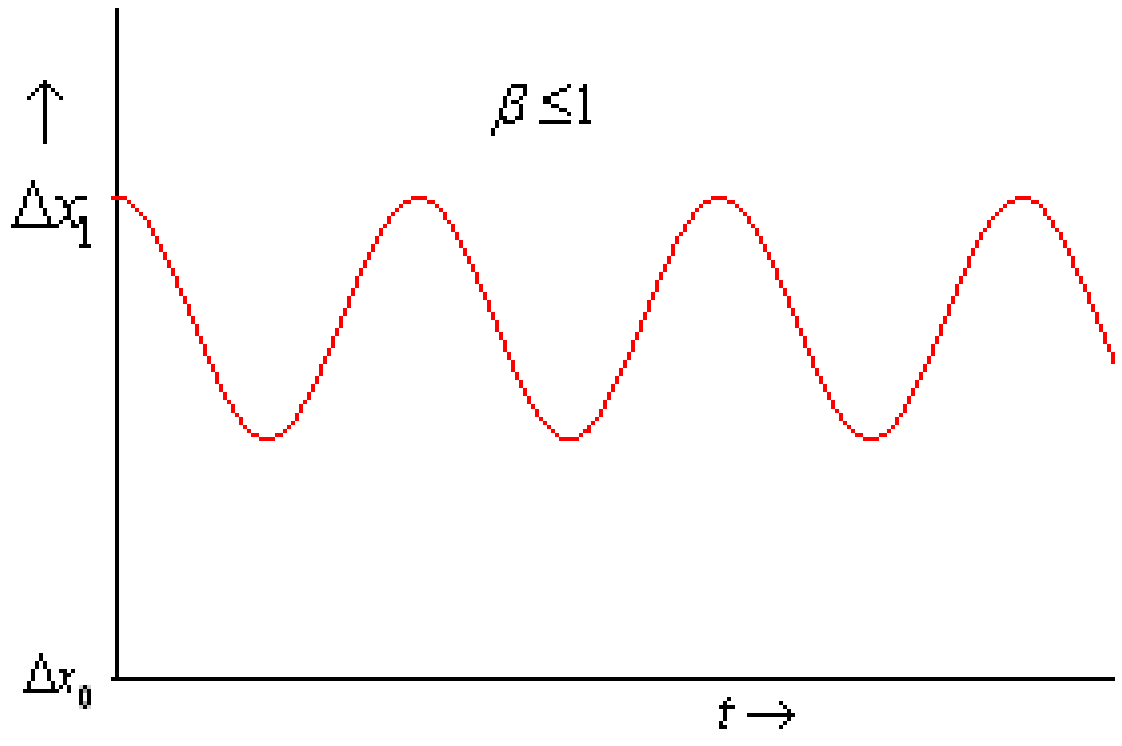

**Fig: 3** Variation of wave-packet width with time for $\beta \leq 1$ as given by eqn. (5.8). As evident there would be no x-quadrature squeezing.

Thus only for $\beta > 1$ **i.e.** $|\alpha|^2$ **(amplitude of oscillation) the width of wave packet decreases below even the ground state width** as shown by figure 2. This behavior is similar to that of **squeezed states** [9, 10]. Also we have,

$$(\Delta x.\Delta p)_1 = \frac{\hbar}{2(1+\beta)}\left[\frac{(\beta^2+3)(\beta+3)}{\beta+1} + \frac{4\beta^2}{(\beta+1)^2}\sin 2(\omega t-\phi)\right]^{\frac{1}{2}} \tag{5.9}$$

We can show very easily that $(\Delta x.\Delta p)_1$ will always be greater than $\frac{\hbar}{2}$ whatsoever be value of $\beta$ be chosen .But uncertainty product is again varying sinusoidally with time.

## 6. Application

Since Squeezed States allow us to evade fundamental Quantum limit set by Uncertainty Principle so in recent times phase squeezing is viewed as a promising tool [11, 12] to detect gravitational radiation [13, 14] ,as predicted by General Theory of Relativity. Single photon excited states have been generated experimentally by Zavatta et al. [15] with good overall efficiency of about 60%. This also marks a beginning for its promising application in Quantum information systems [16] and for possible future applications in the engineering of quantum states [17].

## 7. Conclusion

The influence of fluctuations on Quantum-mechanical oscillator in coherent state can be represented by Excited Coherent States (ECS) which is effectively demonstrated above by ECS wave function. We have also proved that Coherent State can be understood as ground state (i.e. m=0) of the more general Excited Coherent States.

Squeezing of x Quadrature is dependent both on order of excitation (m) and amplitude of oscillation (β). Squeezing is possible only for some values of β precisely for β>1 and as we increase β squeezing decreases however if we increases order of excitation i.e. m then Quadrature squeezing also increases .Thus intensity of the noise (uncertainty in amplitude and phase) can be controlled by varying parameters β and m.

## APPENDIX

To prove :- $\sum_{n=0}^{\infty} \frac{s^n H_{n+m}(\varepsilon)}{n!} = \sum_{n=0}^{\infty} \frac{s^n H_n(\varepsilon)}{n!} H_m(\varepsilon - s)$ ,where $s = \frac{\varepsilon_0}{2}$

Now, we know, $H_n(\varepsilon) = (-1)^n \exp(\varepsilon^2) \frac{d^n}{d\varepsilon^n}\left(\exp(-\varepsilon^2)\right)$

so, $H_m(\varepsilon - s) = (-1)^m \exp\left((\varepsilon - s)^2\right) \frac{d^m}{d\varepsilon^m}\left(\exp\left((\varepsilon - s)^2\right)\right)$

$$= (-1)^m \exp\left(\varepsilon^2 + s^2 - 2\varepsilon s\right) \frac{d^m}{d\varepsilon^m}\left(\exp\left(-\varepsilon^2 + 2\varepsilon s\right)\exp\left(-s^2\right)\right)$$

$$= (-1)^m \exp\left(\varepsilon^2 - 2\varepsilon s\right) \frac{d^m}{d\varepsilon^m}\left(\exp\left(-\varepsilon^2\right)\exp\left(2\varepsilon s\right)\right) \qquad (1)$$

Now applying leibinitz theorem

$$D^m\left(\exp(-\varepsilon^2)\exp(2\varepsilon s)\right) = D^m\left(\exp(-\varepsilon^2)\right)\exp(2\varepsilon s) + {}^m C_1 D^{m-1}\left(\exp(-\varepsilon^2)\right) D^1\left(\exp(2\varepsilon s)\right)$$
$$+ {}^m C_2 D^{m-2}\left(\exp(-\varepsilon^2)\right) D^2\left(\exp(2\varepsilon s)\right) + \ldots\ldots$$

Also, we know $D^n.\exp(mx) = m^n.\exp(mx)$ and $D^n \exp(-\varepsilon^2) = (-1)^n.\exp(-\varepsilon^2).H_n(\varepsilon)$

So, $D^m\left(\exp(-\varepsilon^2)\exp(2\varepsilon s)\right) = (-1)^m \exp(-\varepsilon^2) H_m(\varepsilon)\exp(2\varepsilon s)$

$$+ {}^m C_1 (-1)^{m-1} \exp(-\varepsilon^2) H_{m-1}(\varepsilon).(2s).\exp(2\varepsilon s)$$
$$+ {}^m C_2 (-1)^{m-2} \exp(-\varepsilon^2) H_{m-2}(\varepsilon).(2s)^2.\exp(2\varepsilon s) + \ldots\ldots$$

$$= \exp(-\varepsilon^2)\exp(2\varepsilon s)\begin{bmatrix} (-1)^m H_m(\varepsilon) + {}^m C_1 (-1)^{m-1} H_{m-1}(\varepsilon)(2s) \\ + {}^m C_2 (-1)^{m-2} H_{m-2}(\varepsilon).(2s)^2 + \ldots \end{bmatrix}$$

Now substituting $D^m\left(\exp(-\varepsilon^2)\exp(2\varepsilon s)\right)$ in (1) , we get

$$H_m(\varepsilon - s) = H_m(\varepsilon) + {}^m C_1 (-1)^1 H_{m-1}(\varepsilon)(2s) + {}^m C_2 (-1)^2 H_{m-2}(\varepsilon).(2s)^2 + \ldots$$
$$= H_m(\varepsilon) - {}^m C_1 H_{m-1}(\varepsilon)(2s) + {}^m C_2 H_{m-2}(\varepsilon).(2s)^2 - \ldots\ldots$$

Now R.H.S $= \sum_{n=0}^{\infty} \frac{s^n H_n(\varepsilon)}{n!} H_m(\varepsilon - s)$

$$=\left(1+sH_1+s^2\frac{H_2}{2!}+s^3\frac{H_3}{3!}+\dots\right)\left(H_m(\varepsilon)-{}^mC_1H_{m-1}(\varepsilon)(2s)+{}^mC_2H_{m-2}(\varepsilon).(2s)^2-\dots..\right)$$

Multiplying above terms and rearranging keeping order in mind, we get

$$=\left(H_m+sH_1H_m+s^2\frac{H_2H_m}{2!}+s^3\frac{H_2H_m}{3!}+\dots...\right)$$
$$+\begin{pmatrix}-2s\,{}^mC_1H_{m-1}-2s^2H_1{}^mC_1H_{m-1}-2s^3\frac{H_2}{2!}\,{}^mC_1H_{m-1}\\ +(2s)^2\,{}^mC_2H_{m-2}-sH_1(2s)^2\,{}^mC_2H_{m-2}+\dots\end{pmatrix}$$

where $H_m(\varepsilon)$ is written as $H_m$

$$=H_m+s\left[H_1H_m-2\,{}^mC_1H_{m-1}\right]+\frac{s^2}{2!}\left[H_2H_m-4H_1H_{m-1}{}^mC_1+8H_{m-2}{}^mC_2\right]+\dots....(B)$$

Now, from ghatak[4] we know , $H_{n+1}=(2\varepsilon)H_n-2nH_{n-1}=H_1H_n-2nH_{n-1}$ (2)
where $H_1=2\varepsilon$

so, $H_1H_m-2\,{}^mC_1H_{m-1}=H_{m+1}$

Also, $H_2H_m-4H_1H_{m-1}{}^mC_1+8H_{m-2}{}^mC_2=H_2H_m-4mH_1H_{m-1}+4m(m-1)H_{m-2}$

$$=\left(H_1^{\,2}-2\right)H_m-4mH_1H_{m-1}+4m(m-1)H_{m-2}$$

(substituting $H_2$ using (2) )

$$=H_1\left(H_1H_m-4mH_{m-1}\right)-2H_m+4m(m-1)H_{m-2}$$
$$=H_1\left(H_{m+1}-2mH_{m-1}\right)-2H_m+4m(m-1)H_{m-2}$$

(again using(2)

$$=H_1H_{m+1}-2H_m-2mH_1H_{m-1}+4m(m-1)H_{m-2}$$
$$=H_1H_{m+1}-2H_m-2m\left[H_1H_{m-1}-2(m-1)H_{m-2}\right]$$

Now Replacing n by (m-1) in (2), we get $\Rightarrow H_1H_{m-1}-2(m-1)H_{m-2}=H_m$

$$= H_1 H_{m+1} - 2H_m - 2mH_m = H_1 H_{m+1} - 2\left(m+2\right) H_m$$

$$= H_{m+2} \qquad \left(\because \text{ Replacing n by m+1 in(2)and Applying the result}\right)$$

So finally (B) becomes

$$= H_m + sH_{m+1} + \frac{s^2}{2!} H_{m+2} + \ldots\ldots..$$

$$= \sum_{n=0}^{\infty} \frac{s^n H_{n+m}\left(\varepsilon\right)}{n!} = \text{L.H.S}$$

Hence Proved !